\documentclass[submit]{aiaa-tc}

\usepackage{amsmath,graphicx,amssymb}

 \usepackage{varioref}
 \usepackage{wrapfig}
 \usepackage{threeparttable}
 \usepackage{dcolumn}
  \newcolumntype{d}{D{.}{.}{-1}}
 \usepackage{nomencl}
  \makeglossary
 \usepackage{subfigure}
 \usepackage{subfigmat}
 \usepackage{fancyvrb}
  \fvset{fontsize=\footnotesize,xleftmargin=2em}
 \usepackage{lettrine}
 \usepackage[colorlinks]{hyperref}

 \title{The Role of Data Analysis in Uncertainty Quantification: \\ Case Studies for Materials Modeling}

 \author{
  Paul N. Patrone  \thanks{Applied and Computational Mathematics Division, National Institute of Standards and Technology, 100 Bureau Drive, Gaithersburg MD, 20899, USA.}, Anthony J. Kearsley \thanksibid{1},
  Andrew M. Dienstfrey  \thanksibid{1}
 }

 \def\E{\boldsymbol {\rm E}}
\def\bG{\boldsymbol {\rm G}}
\def\bsig{\boldsymbol {\rm \Sigma}}
\def\mN{\boldsymbol {\mathcal N}}
\def\e{{\rm e}}

\def\hvs{\hat {\boldsymbol \varsigma}}
\def\bA{\boldsymbol {\rm A}}
\def\bb{\boldsymbol {\rm b}}

\begin{document}

\maketitle

\begin{abstract}
In computational materials science, mechanical properties are typically extracted from simulations by means of analysis routines that seek to mimic their experimental counterparts.  However, simulated data often exhibit uncertainties that can propagate into final predictions in unexpected ways.  Thus, modelers require data analysis tools that (i) address the problems posed by simulated data, and (ii) facilitate uncertainty quantification.  In this manuscript, we discuss three case studies in materials modeling where careful data analysis can be leveraged to address specific instances of these issues. As a unifying theme, we highlight the idea that attention to physical and mathematical constraints surrounding the generation of computational data can significantly enhance its analysis.
\end{abstract}


\lettrine[nindent=0pt]{I}{ncreasingly}, the computational materials science community is embracing uncertainty quantification (UQ) as a necessary component of any modeling workflow that aims to provide actionable information for industry.\cite{ICME1,IMAJOM,economics,ASME}  In many instances, this change has been driven by the observation that computational predictions, while less expensive than their experimental counterparts, are affected by uncertainties not typically encountered in laboratory settings.  As a result, there has been significant work focused on identifying and quantifying uncertainties associated with simulation tools {\it per se}.\cite{CalVal1,Rizzi1,Rizzi2}  Despite this, however, much less effort has been devoted to understanding the impact of these uncertainties on subsequent analyses and usage models.  Critically, such issues must be addressed if the benefits of computational approaches are to be fully realized.

Within this greater context, data analysis remains a fundamental and sometimes overlooked task that can contribute to uncertainties in ways that are difficult to quantify, if not control.  For example, many simulated properties are determined by means of analysis routines that have direct analogs to real experiments, e.g.\ as in the estimation of yield strain from a virtual stress-strain curve.\cite{Patrone16,Patrone17}  Oftentimes, however, such methods are taken directly from the experimental realm without consideration as to whether they require modification or are even applicable.  In some cases, experience has even shown that these analyses become unstable when subject to typical uncertainties that arise from standard modeling tools.\cite{Patrone16}

In such light, this manuscript considers several case studies in materials modeling where careful data analysis can be leveraged to better understand and quantify uncertainties that arise within the context of simulations and theoretical models.  We emphasize that our goal is not so much to provide universal recipes on how to analyze data, but rather to show how the quality of estimates can be improved though attention to the physical and mathematical constraints underlying data.   Moreover, we discuss how this level of attention allows one to quantify uncertainties associated with the structural choices of the analysis itself, an issue that must be addressed as part of any complete uncertainty budget.

For definiteness, we consider data obtained from molecular dynamics (MD) simulations of condensed matter systems.  In the past few decades, MD has become popular in the aerospace community for its ability to model the complicated and heterogeneous structure of thermoset polymers used in composites.\cite{Strachanreview,Varshney08,Yu09,stevepat1,stevepat2,Khare09,Fan07,Khare12,Khare13} While a complete review of MD is far beyond our scope, it is important to note that this technique is typically limited to modeling $\mathcal O(10^4)$ atoms, corresponding to length and time scales on the order of nanometers and nanoseconds.  An immediate consequence of this limitation is that MD data  suffers exaggerated effects of thermal noise and discrete-particle motion, both of which hinder precise data analysis.  Perhaps worse, the small system-sizes call into question the validity of simulated results as a proxy for bulk data.  In the following, we therefore consider ways in which analysis routines can mitigate some of these issues.  It is also worth emphasizing that while our main focus is on MD simulations, many of the approaches considered here are applicable to a wider range of modeling techniques.  

A central theme that unifies our discussion is the general observation that  analysis methods often perform better when they leverage the global behavior of data.  For example, material properties such as the glass-transition temperature or yield strain are identified in terms of critical-points (i.e.\ a local property) of corresponding modulus and stress-strain curves,\cite{Donth,Patrone16,Patrone17} which we denote generically as $f(x)$.  In physical experiments, this approach is usually sound, given that the corresponding curves are relatively noise-free and easy to interpret.  In simulated data, however, this is often not the case, so that, e.g., we can only locate critical points $x_c$ to within a factor of $\sqrt{\sigma / f''(x_c)}$, where $\sigma$ is the characteristic noise scale (which is large) and $f''(x_c)$ is the second derivative of the signal at $x_c$.  Oftentimes, a global analysis can mitigate this problem by using adjacent, {\it non-extremal} data along with modeling assumptions to better localize critical points.  Importantly, these considerations are not restricted to standard curve-fitting, but can also encompass more general approaches such as convex optimization, which extends to situations where an admissible fit-function is unknown.  Moreover, considerations of global behavior are not restricted to the task of estimating critical points but also pertain to other analyses as well.  To illustrate these issues, we therefore consider three examples of data analysis that are representative of tasks routinely performed in industrial settings.  

In the first example, we consider the most basic task of averaging time-evolving observables computed via MD.  In particular, we challenge the common claim that each realization of the observable should be independent, which has led to the widespread practice of discarding the majority (often $99\%$ or more) of data computed by a simulation.  Rather, we show that by accounting for autocorrelation of the simulated measurements, it is in fact possible to faithfully compute ensemble averages {\it and even reduce uncertainties, despite lack of independence in the data}.  

In the second example, we consider the commonly encountered task of inferring the onset of a transition (e.g.\ the glass transition, yield, etc.) in terms of bilinear behavior.\cite{Patrone16,Patrone17}  A typical approach for analyzing such data amounts to fitting two lines to ``asymptotic'' regimes and computing their intersection.  However, it is straightforward to show that this approach suffers from a ``lever-arm'' effect that amplifies uncertainties through extrapolation.  As an alternative, we therefore discuss global fits in terms of functions such as hyperbolas that naturally identify asymptotic regimes without the need for a user to subjectively identify them.  Critically, this approach allows one to automatically assess the quality of data by directly checking that it has the necessary characteristics for estimating the transition.

In the third example, we extend this idea to data for which there is no known generic fit-function, namely simulated stress-strain curves.  Historically, these have been used to determine the yield-strain of a material, typically in terms of the ordinate of the global maximum.  However, estimating this quantity from MD simulations is often difficult, especially in high-throughput settings where system sizes are small, and thus noise is large.  In this context, we consider the simple observation that a stress-strain curve is concave before onset of strain-hardening, which can be leveraged to provide reasonable estimates of yield.  Moreover, the associated convex optimization algorithms indicate when the data is statistically non-concave, thereby offering a degree of self-assessment analogous to our second example.


It is worth emphasizing that we draw most of our examples from high-throughput simulations that model small systems [i.e. $\mathcal O(10^3)$ atoms] compared with what is achievable by today's machines [e.g. $\mathcal O(10^6)$ atoms or more].  While it is generally preferable to simulate the largest system possible, such computations can take on the order of weeks or months, which is infeasible in many industrial settings.  Thus, we feel that the data and issues presented here are more representative of those facing our readers.  We also mention that this manuscript is primarily concerned with the impact of data analysis on uncertainty quantification, not UQ {\it per se}.  As such, we assume some level of familiarity with tools such as noise modeling, parametric bootstrap analyses, and synthetic dataset generation.\cite{smith2013,RW,hesterberg:2011,Calibration1,ferson}  In a few instances we review key ideas, but interested readers should consult the indicated references for detailed information.  

\section{Correlations are not scary}

\begin{figure}
\includegraphics[width=16cm]{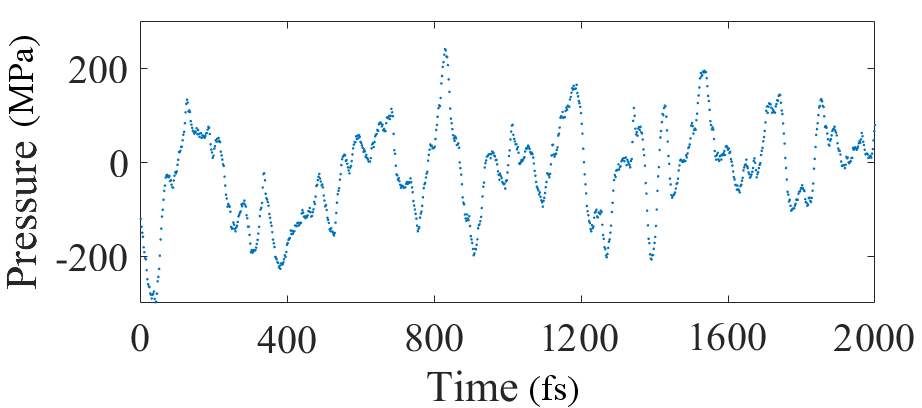}\caption{Pressures computed for a coarse-grained water simulation with 125 molecules.  Note that the characteristic scale of noise is 100 MPa, roughly three orders of magnitude larger than the mean pressure (which is on the order of 100 KPa).  Also, pressures are correlated over the scale of hundreds of fs.}\label{fig1}
\includegraphics[width=16cm]{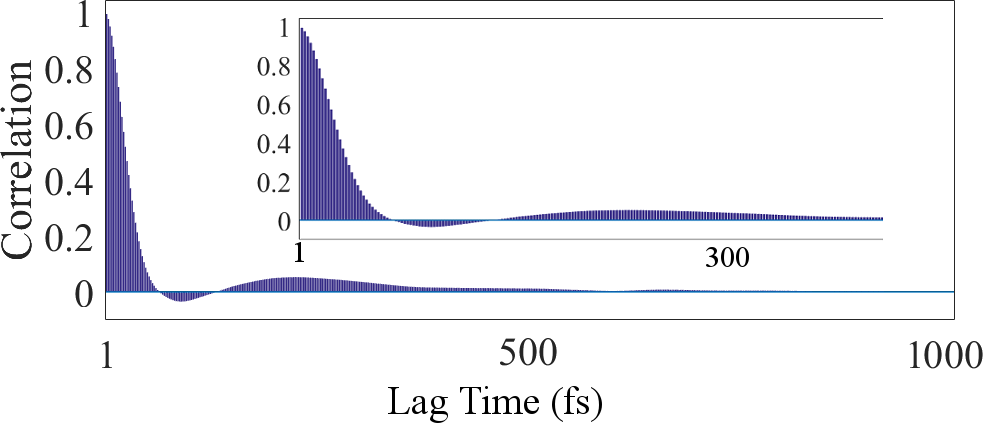}\caption{Pressure autocorrelation estimated from $10^7$ timesteps for the same system as in figure \ref{fig1}.  Note that the correlations fall to approximately zero after roughly 600 fs, which corresponds to 300 timesteps.  The inset shows a closer view of the first 400 fs.}\label{fig2}
\end{figure}

Molecular dynamics (and many related simulation methods) compute material properties in terms of ensemble averages.\cite{allen89,binder95}  Generally speaking, if $G(x)$ is the microscopic counterpart of an observable $\langle G \rangle$ (e.g.\ pressure, density, etc.) given in terms of microstate $x$ (i.e.\ particle positions and momenta), then the expected value $\langle G \rangle$ is given by the arithmetic mean
\begin{equation}
\langle G \rangle_N = \frac{1}{N}\sum_{k=1}^N G(x_k) \label{eq:sampmean}
\end{equation}
where $x_k$ is a time series of microstates generated by the simulation algorithm and $\langle \star \rangle_N$ refers to a sample average according to Eq.~\eqref{eq:sampmean}.  Statistical mechanics makes the further assumption that over long times, the probabilities of various microstates are described by a density $P(x)$ that is a function of the system energy.\cite{Pathria}  But because the $x_k$ are generated as a time-series,  the linear correlation function is typically non-zero, i.e.\
\begin{equation}
\E\left\{ [G(x_j) - \E(G) ][G(x_k) - \E(G) ] \right \} \ne 0, \label{eq:correlation}
\end{equation}
where the expectation $\E(G) $ is with respect to the joint distribution of $x_k$ and $x_j$.  In effect, Eq.\ \eqref{eq:correlation} states that the system retains some memory of its past.   For systems in equilibrium, however, the properties of the system are time-invariant.  This implies that the correlation function given by Eq.~\eqref{eq:correlation} only depends on the difference of $j$ and $k$, namely, 
\begin{equation}
\E\left \{ [G(x_j) - \E(G)][G(x_k) - \E(G)] \right \} = \sigma^2C(|j-k|), \label{eq:stationary_correlation}
\end{equation}
where $C$ is an undetermined function.\cite{SalI,SalII}  Moreover, it is generally observed that $C\to 0 $ as $|j-k| \to \infty$; that is, the system eventually forgets the distant past.  See, for example, Figs.~\ref{fig1} and \ref{fig2}.

Motivated by this observation, modelers frequently conclude that the safest way to construct estimates via Eq.~\eqref{eq:sampmean} is to use a subset of independent $x_k$ taken from the available data.  That is, one determines a {\it correlation time} $\tau$ beyond which $C(t/ \Delta t>\tau) \approx 0$ where $\Delta t$ is the fixed time-step of the MD integrator.\cite{SalI,SalII}  Given a series of microstates $x_0,x_1,...$, one then replaces Eq.~\eqref{eq:sampmean} with the expression
\begin{equation}
\langle G \rangle_n = \frac{1}{n}\sum_{k=1}^nG(x_{k\tau}),\label{eq:subsetmean}
\end{equation}
where $n=\lfloor N/\tau \rfloor$ is the number of independent samples, and $\lfloor z \rfloor$ denotes the greatest integer less than $z$.  Approximating the variance $\sigma^2$ in terms of the experimental standard variance (i.e.\ the sample variance)
\begin{equation}
\varsigma^2 = \frac{1}{n-1}\sum_k \left(G(x_{k\tau}) - \langle G\rangle_n \right)^2, \label{eq:uncorrstd}
\end{equation}
one can estimate the uncertainty in $\langle G \rangle_n$ relative to $\E(G)$ (i.e.\ the standard variance of the mean) by $\epsilon_\varsigma^2 = \varsigma^2/n$.

From a statistical perspective, the intuition behind Eqs.~\eqref{eq:subsetmean} and \eqref{eq:uncorrstd} amounts to the observation that correlated data does not add new information about the expected value $\E(G)$ when used in Eq.~\eqref{eq:sampmean}.  In general, however, this intuition is misguided.  The time-series $x_k=(-1)^k$, for example, has a mean value $\langle x \rangle_N \to 0 = \E(x)$ as $N\to \infty$, and all pairs of $x_j,x_k$ are either perfectly correlated or anticorrelated.  In principle, one could generate an uncorrelated subset of $x_k$ by drawing $n \ll N$ samples at random from the original set, but one does much better (in terms of reducing $\varsigma^2$) by keeping all of the data because the ``errors'' in each pair of $x_k,x_{k+1}$ cancel.  More generally, negative correlations can reduce uncertainties, while correlations less than unity still provide useful information.

To see this in more detail, consider the situation described by Eq.~\eqref{eq:stationary_correlation} in which the correlations in $G_j,G_k$ depend only on the relative difference of indices $|j-k|$.  To make a connection with MD, we also assume that $N\gtrsim \mathcal O(10^6)$ and $\tau \ll N$, which is quite reasonable (if not an understatement) for most implementations. For simplicity, we further assume that the the $G_j$ are normally distributed, which leads to the following model for $\bG=(G_1,...,G_N)$:
\begin{align}
\bG = \E(\bG) + \mN(0,\bsig).
\end{align}
In this expression, $\mN$ is a Gaussian random vector with zero mean and covariance matrix $\bsig$ given by
\begin{align}
\bsig = \sigma^2
\begin{bmatrix}
1 & C_1 & \dots & C_{N-1} & C_N  \\
C_1 & 1 & \ddots & C_{N-2} & C_{N-1} \\
\vdots & \ddots & \ddots & \ddots & \vdots \\
C_{N-1} & C_{N-2} & \ddots & 1 & C_1 \\
C_N & C_{N-1} & \dots & C_1 & 1
\end{bmatrix}, \label{eq:corrmat}
\end{align} 
where $C_k=C(k)$ is the stationary correlation evaluated for the indicated difference of indices.  Mathematically, the probability of realizing a particular collection $\bG$ is given by the density
\begin{align}
P(\bG) = \frac{1}{\sqrt{2\pi |\bsig|}} \exp\left[-(1/2)\bG^{\rm T} \bsig^{-1} \bG \right],
\end{align} 
where $|\bsig|$ is the determinant of $\bsig$.

Several comments are in order.  It is easy to show, for example, that the sample mean $\langle G \rangle_N$ using the correlated values is still an unbiased estimator of $\E(G)$.  In other words,
\begin{align}
\E[\langle G \rangle_N] = \E(G) 
\end{align}
by virtue of the fact that the noise has zero mean.  Correlations between the individual $G_k$ play no role in this conclusion.  Second, it is straightforward to compute the variance of the mean.  In particular, note that
\begin{align}
\epsilon_\sigma^2 = \E\left\{ [\langle G \rangle_N - \E(G)]^2 \right\} &= \frac{1}{N^2}\sum_{j,k} \E\left\{ [G_j - \E(G))][G_k - \E(G)] \right\} \nonumber \\
& = \frac{\sigma^2}{N^2}\sum_{j,k} C_{|j-k|} = \frac{\sigma^2}{N} + \frac{\sigma^2}{N^2}\sum_{\substack{j,k\\j\ne k}}C_{|j-k|}  \label{eq:correrror} \\
&\approx \frac{\sigma^2}{N} + \frac{2\sigma^2}{N}\sum_{j=1}^NC_{j}, \label{eq:approxerror}
\end{align}
where the last approximation requires that $\tau \ll N$ and ignores ``boundary effects'' associated with data at the beginning and end of the time series.\footnote{The Toeplitz structure of the covariance matrix is the discrete analog of the stationarity property.  Equation \eqref{eq:approxerror} arises by adding non-zero terms to the lower-left and upper-right corners of the covariance matrix in order to make it circulant.  Physically, this corresponds to assuming that the first timesteps are correlated with the last, and vice versa.  While untrue, the error in this approximation is negligible when $\tau \ll N$.}  In the case of uncorrelated data (that is, $C_{|j-k|} = \delta_{j,k}$, where the latter is the Kronecker delta), then Eq.~\eqref{eq:correrror} becomes the usual result that $\E\left\{ [\langle G \rangle_N - \E(G)]^2 \right\} = \sigma^2/N$, for which $\epsilon_\varsigma^2=\varsigma^2/N$ is often a reasonable estimate.  However, Eq.~\eqref{eq:correrror} is more general in that it applies to arbitrary covariance matrices, provided they are stationary.  The corresponding $\varsigma^2$ can be computed from the usual formula in Eq.~\eqref{eq:uncorrstd}, but using all of the available data.\footnote{That is, Eq.~\eqref{eq:uncorrstd} amounts to an estimate of $\bsig_{1,1} = \sigma^2$.  Technically speaking, this is estimate is only asymptotically unbiased, meaning that Eq.~\eqref{eq:uncorrstd} exhibits a small bias that is $\mathcal O(\tau/N)$.  Given, however, that $\tau=\mathcal O(10^2)$ and $N=\mathcal O(10^6)$ are typical numbers encountered in simulated data, this bias is often negligible.}  The covariance matrix elements $C_k$ can likewise be estimated via an autocorrelation analysis (available in many software packages) on the data.\footnote{The considerations of the previous footnote also apply to the covariance calculated in such a manner.  Care should always be taken to ensure that the ratio $\tau/N\ll 1$.}  

A bit of numerology illustrates the usefulness of Eq.~\eqref{eq:correrror}.  In particular, if we recall our assumption that the correlation time $\tau \ll N$ (which is typical of simulated data), then $\sum_{j,k}C_{|j-k|}$ has on the order of $N\tau$ non-zero elements.  As these are all bounded from above by $1$, we immediately see that $\epsilon_\sigma^2 \lesssim \sigma^2 \tau /N = \epsilon_\varsigma^2$.
Note the appearance of $n=N/\tau$, which corresponds to the number of uncorrelated samples that arise from using subsets of the data [cf.~\eqref{eq:uncorrstd}].  At first glance, this inequality does not seem to support the idea of using the correlated data, since we may only obtain the same level of accuracy using an uncorrelated subset.  However, in many physical problems $C_j$ decays according to some power law or exponential, so that our inequality drastically overestimates the contribution from $C_{j>0}$.  Moreover, physical systems tend to exhibit a certain degree of anticorrelation over time, so that [as in the example of $x_k=(-1)^k$] Eq.~\eqref{eq:correrror} benefits from cancellation effects that reduce uncertainty.  

\begin{figure}
\includegraphics[width=16cm]{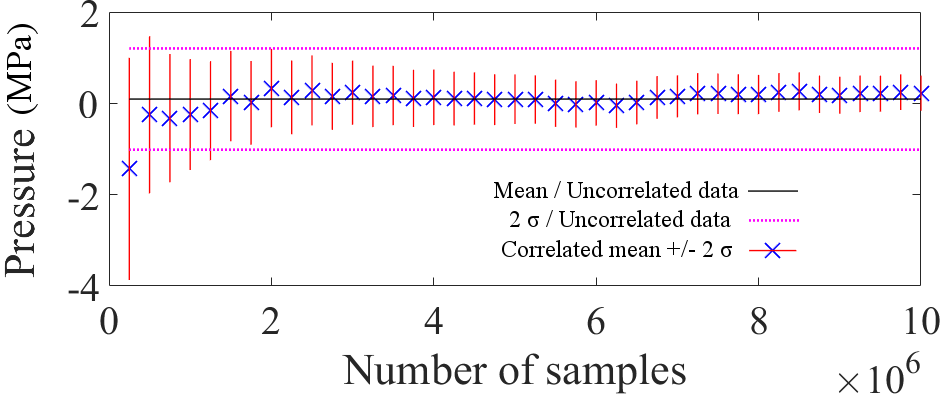}\caption{Comparison between the standard error computed from sparse but uncorrelated subsampling of $10^7$ simulated pressures (horizontal lines) and estimates computed according to Eq.~\eqref{eq:approxerror} (blue $\times$ with vertical error bars).  For the latter, the horizontal axis indicates the number of sequential timesteps used to compute the estimates.  Note that using less than $1/5$ of the simulated data, the correlated estimates are statistically equivalent to their uncorrelated counterparts.  Using more of the available data, the correlated estimates have corresponding less uncertainty.  This suggests that we waste information by subsampling data to generate uncorrelated samples.}\label{fig3}
\end{figure}   

To better illustrate this issue, we  consider the task computing the expected pressure of a coarse-grained water simulation using MD.\cite{Hoover,LAMMPS,MATLAB}\footnote{Details of the simulation are as follows.  The system was composed of 125 point particles (each representing a water molecule) at a number density of $33.6$ nm${}^{-3}$.  Temperature control was enforced by the Nos{\' e}-Hoover thermostat at $300$ K.  Numerical integration was done using LAMMPS.  The system was equilibrated for 100,000 timesteps of 2 fs each.  After this, we ran a production run of 10,000,000 timesteps, outputting pressures at each iteration.  The force-field (which is available upon request, along with other simulation files) was a custom, tabulated function having a range of $0.75$ nm.  It was calibrated through a gradient-descent algorithm (unpublished).  The latter attempts to match the coarse-grained radial-distribution function and pressure with a corresponding atomistic simulation.  Analysis was done using custom MATLAB scripts.}\footnote{Certain commercial products are identified in this chapter in order to specify the computational procedure adequately. Such identification is not intended to imply recommendation or endorsement by the National Institute of Standards and Technology, nor is it intended to
imply that the materials or equipment identified are necessarily the best available for the purpose.}  In practice, this computation often comes with large uncertainties, especially for small systems; cf.\ figure \ref{fig1}.  In order to avoid correlations, common practice is to keep only every 100th or 1000th timestep, which amounts to discarding 99\% or more of the data!  To compare this approach with Eq.~\eqref{eq:approxerror}, we therefore construct a mean and experimental standard uncertainty of the mean (i.e.\ the standard error) by keeping every $300$th pressure (cf.\ figure \ref{fig2}).  Figure \ref{fig3} shows these values relative to estimates constructed using all of the correlated data, with uncertainties computed according to Eq.~\eqref{eq:approxerror}.  Notably, estimates according to Eq.~\eqref{eq:approxerror} are statistically equivalent to their uncorrelated counterparts using data from only the first 20\% of the simulation.  In other words, the uncorrelated subsample estimate requires five times more data to generate estimates that are as good as those given by Eq.~\eqref{eq:approxerror}. 

We emphasize that the exact level of computational savings associated with Eq.~\eqref{eq:correrror} will vary by application.  Nonetheless, the crux of our argument still holds: one can do no worse by averaging correlated data, and often one stands to gain significantly.  From the standpoint of reproducibility, however, there are other advantages that, in our minds, justify the added effort.  For one, experience has shown that it is easy to fall into the trap of relying on rules of thumb (e.g.\ always saving data from the $100$th timestep) without testing the assumptions underlying those rules.  Moreover, it can be difficult to assess that a simulation is outputting data consistent with thermal equilibrium.  By forcing users to be more engaged with the correlation structure and properties of their data, our approach therefore allows one to better assess its quality to ensure that it is physically meaningful.  In light of the large costs of many modeling protocols, such assessments are necessary for building confidence in simulated results.

 
\section{Getting the most out of your (bilinear) fit}

\begin{figure}\begin{center}
\includegraphics[width=12cm]{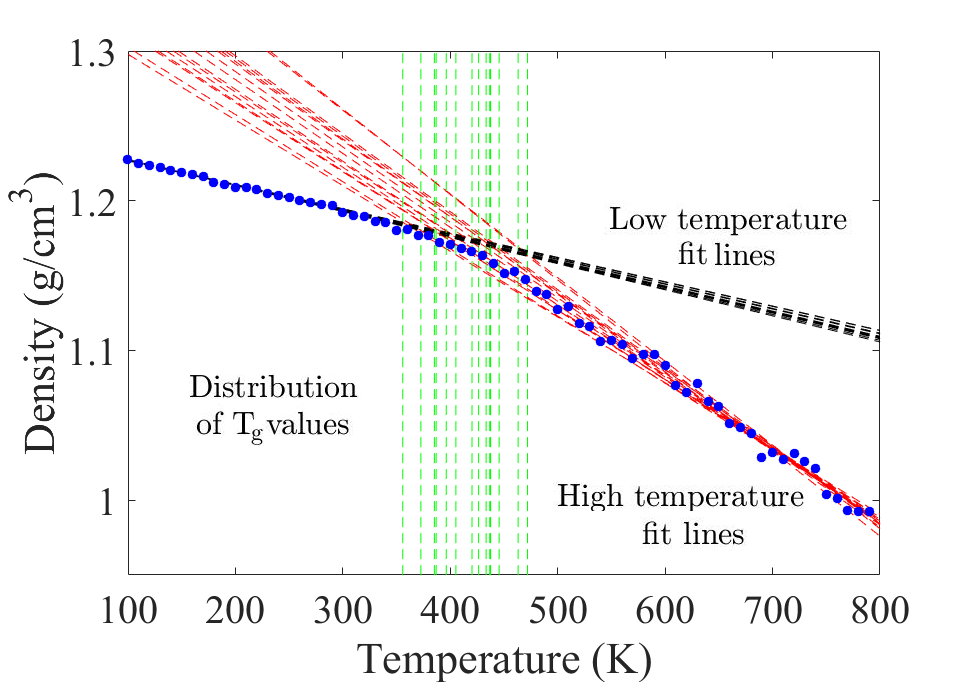}\caption{Example of bilinear data extracted from a MD simulation of the glass-transition.  The low and high-temperature lines are fit to the first and last 15 data-points of synthetic datasets generated from a hyperbola regression of the original data; see  Ref.~\citen{Patrone16} for more details of the associated modeling.  Note that the dispersion in high-temperature fit lines increases significantly as one extrapolates away from the high-temperature regime.} \label{fig:hyperbola}
\includegraphics[width=12cm]{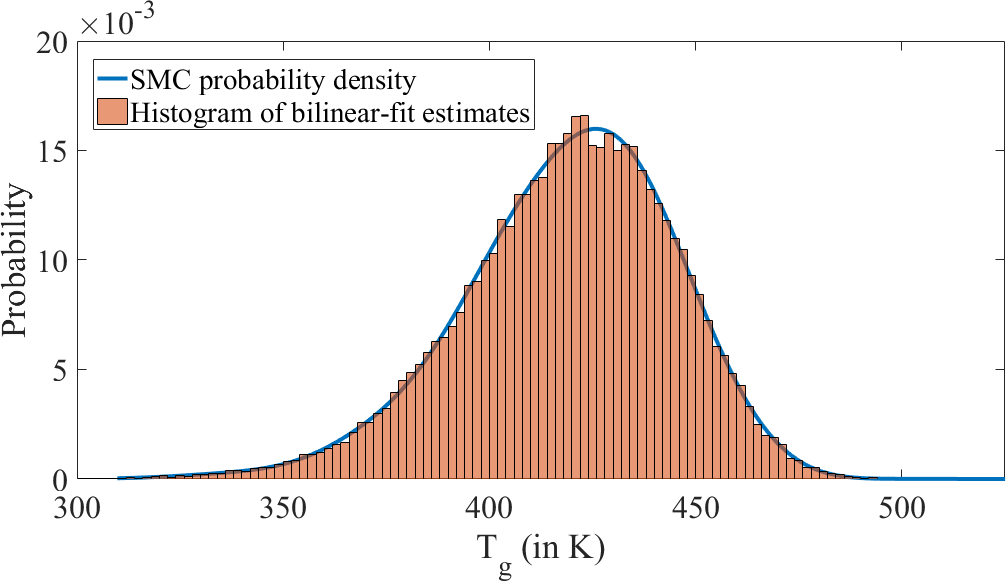}\caption{Spread in $T_g$ value computed according to the method in figure \ref{fig:hyperbola} when iterated 50,000 times.  [The spectral Monte Carlo (SMC) estimate of the probability density is discussed in Ref.~\citen{PatroneRosch17}.]  Note that the range of plausible $T_g$ values associated with this method spans roughly $150$ K, which is too large for practical purposes.}\label{fig:hist}
\end{center}
\end{figure}

Experimental and simulation-based procedures often identify material properties by means of a change in the functional relationship between two quantities.
A common example encountered in materials modeling are analyses of bilinear datasets, i.e.\ ones that exhibits two asymptotically linear regimes ``connected'' by a transition domain; see figure \ref{fig:hyperbola}.  The {\it location} of this transition domain is often associated with the material property of interest, so that the corresponding analysis methods amount to characterizing and pinpointing geometrical structures within the data.

In experimental data, such transition domains are often sharp and noise-free because the length and time-scales of the measurement are large relative to those of the atomistic processes determining the property.\footnote{In other words, bulk-scale experiments describe the ``average'' behavior of a system, which smooths out noise and finite-size effects.}  In MD simulations, however finite-size effects and thermal noise are more pronounced because the system sizes are small [e.g.\ $\mathcal O(10^4)$ atoms] compared to bulk scales [$\mathcal O(10^{23})$ atoms].  As a result, corresponding transitions in the data may occur over essentially the entire domain, so that asymptotic regimes are difficult to identify and analyze.  This leads to a situation in which analysis routines applicable to experimental data are no longer useful for simulations.

Figure.~\ref{fig:hyperbola} illustrates this problem in the context of the glass-transition temperature $T_g$.  Convention dictates that this quantity can be extracted from the intercept of two lines fit to the low and high-temperature regimes of a simulated density-temperature curve.  As the figure shows, however, noisy data leads to several problems.  For one, it is not clear where the high-temperature asymptotic regime begins.  Second, a variety of slopes characterize its behavior (assuming we can identify its location).  Third, extrapolating to low-temperatures {\it amplifies uncertainties} via a lever-arm effect.  To illustrate the impact of these problems, figure \ref{fig:hist} shows a histogram of $T_g$ values computed from a noise-modeling approach discussed in other works.\cite{Patrone16,Patrone17}  Given that the range of plausible $T_g$ values predicted by the bilinear fit spans roughly 150 K, we see that this method may not be practical for simulated data.

Conceptually, a key problem with this type of analysis is that it (i) assumes a certain structure to the data, and (ii) requires that the modeler subjectively identify relevant structural components.  From an UQ standpoint, this can lead to unreasonable predictions and/or uncertainties if either (a) the data does not contain the features required by the analysis, or (b) the method is overly sensitive to the subjective choices of the modeler.  {\it This suggests that we would be better served by fitting routines that not only characterize data, but also interrogate it.}

In the context of bilinear data, this more stringent requirement can be addressed, for example, using hyperbolas.  Notably, such functions transition between two asymptotic, linear regimes in a smooth manner reminiscent of the data in figure \ref{fig:hyperbola}.  Thus, fits to hyperbolas interrogate the global structure of data, allowing one to more objectively decide if it indeed exhibits bilinear behavior.  Moreover, in doing so, the fit automatically identifies the asymptote intersection (i.e.\ the hyperbola center), which is conceptually equivalent to the output of a bilinear fit.  As an added bonus, we avoid the possibility of extrapolating linear behavior and thereby amplifying noise.

As several works have explored such analyses in detail,\cite{Patrone16,Patrone17} we only sketch a few key ideas of the central argument.  From a practical perspective, it is useful to adopt a parameterization of the form
\begin{align}
\rho(T) &= \rho_0 - a (T - T_0) - b \mathcal H_0 (T,T_0,c), \nonumber \\
\mathcal H_0(T,T_0,c) & = \frac{1}{2} (T-T_0) +  \sqrt{\frac{(T-T_0)^2}{4} + \exp(c)},\label{eq:hyperbola}
\end{align}
where $(T_0,\rho_0)$ is the hyperbola center and $a,b,$ and $c$ are constants.  Note that $c$ effectively controls the scale of the transition between asymptotes, which have slopes $-a$ and $-a-b$.  We denote these by $\rho'_{-\infty}$ and $\rho'_{\infty}$ respectively. 
Using Eq.~\eqref{eq:hyperbola}, it is straightforward to calculate the relative convergence of the  hyperbola slope to  either of its asymptotic limits. For the higher asymptote, one finds 
\begin{align}
\mathcal P_h(T) 
  &:= \frac{\rho'(T)-\rho'_{-\infty}}{\rho'_{\infty}-{\rho'_{-\infty}}} \\
  &= \frac{1}{2} + \frac{(T-T_0)}{2 \sqrt{(T-T_0)^2 + 4 \e^{c}}}. \label{eq:pequation}
\end{align}
By symmetry, $\mathcal P_\ell = 1-\mathcal P_h(T)$ gives the percent convergence to the lower asymptote.

\begin{figure}
\includegraphics[width=16cm]{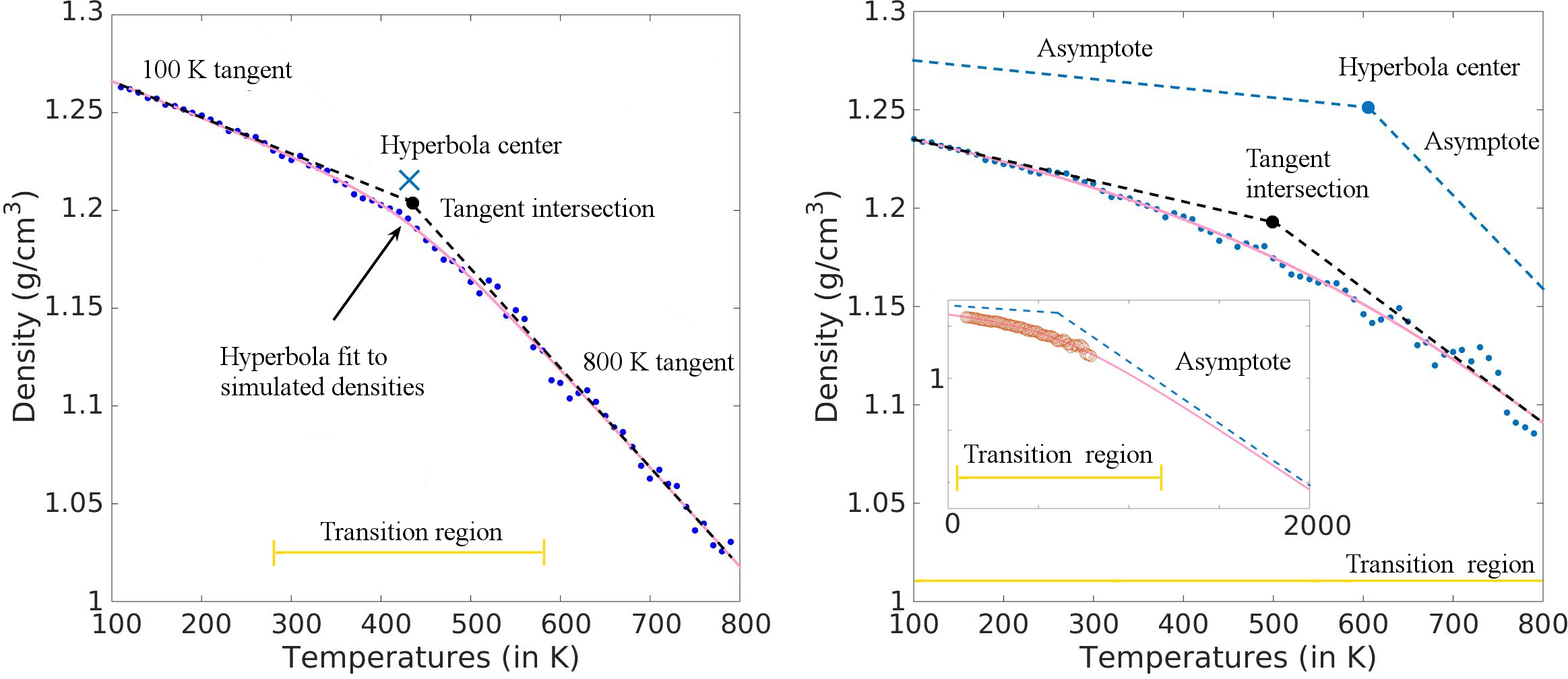}\caption{Hyperbolas fit to different glass-transition simulations run with MD.  The yellow line marked ``Transition region'' indicates the domain of temperatures of which the data is not asymptotic, with $\mathcal Q=90$ \% indicating the level of required convergence.  Because the data entirely contains the transition region, it is reasonable to assume that the simulation samples both low and high-temperature asymptotic behavior.  The subplot on the right shows a dataset for which this is not the case.  }\label{fig:hypcomp}
\end{figure}

With these observations in mind, a hyperbola can be used to assess the quality of data and extract $T_g$ as follows.  First, specify a ${\mathcal Q}$ that characterizes the minimum percentage to which some of the data should be converged to the asymptotes.  Next, fit Eq.~\eqref{eq:hyperbola} to a density-temperature curve and compute $T_h = \mathcal P_h^{-1}(\mathcal Q)$ and $T_\ell=\mathcal P_\ell^{-1}(\mathcal Q)$, where $\mathcal P_\star^{-1}$ refers to the inverse of Eq.~\eqref{eq:pequation} and its low temperature counterpart.  Any dataset for which $T_h$ is greater than the highest simulated temperature (or $T_\ell$ below the lowest) does not sample the corresponding asymptotic regimes to within the desired level of convergence and can be held out for further investigation.  For datasets that do satisfy this criterion, $T_0$ can be taken as the corresponding estimate of $T_g$.  

Figure \ref{fig:hypcomp} illustrates these ideas for two datasets representative of results obtained from simulating the glass-transition of the thermoset systems 3,3- diaminodiphenyl sulfone bisphenyl
F (33DDS-BisF) and a mixture of 3,3- diaminodiphenyl sulfone, 4,4-diaminodiphenyl sul-
fone, and tetraglycidyl methylene dianiline (3344MY); see Refs.~\citen{Patrone16} and \citen{Patrone17} for details of these systems.   In these examples, we set $\mathcal Q=90$ \% as the level of convergence required for the dataset to adequately sample asymptotic behavior.  The left plot shows a system for which the corresponding transition region is contained entirely within the domain of the data.  The corresponding low and high-temperature data points have a relatively easy to identify linear regime (although this is done automatically by the fit).  The right plot, by contrast, shows a system for which the asymptotic behavior is difficult to identify.  This conclusion is encapsulated in the width of the transition region, which extends far beyond the domain of the data.  This dataset in particular is a candidate for closer inspection and/or rejection on the basis that it is not representative of a physically meaningful glass-transition.

\section{When there is no model functional form}

\begin{figure}
\includegraphics[width=16cm]{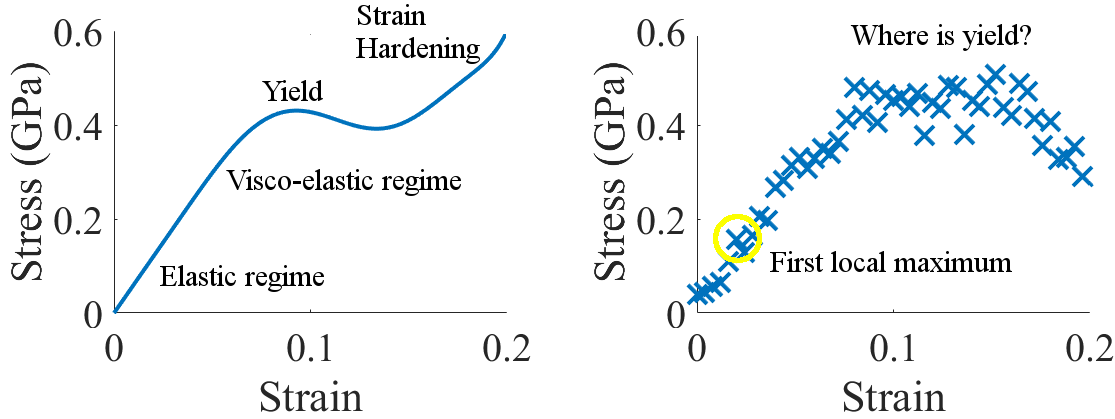}\caption{Comparison of an idealized stress-strain curve and its simulated counterpart.  Left: In experimental data, such curves often show four distinct regimes characterizing: elastic behavior; visco-elastic behavior; yield; and strain hardening.  Generally speaking, structural components (such as wings) tend to fail when the material reaches yield, which is often identified as the first maximum of a stress-strain curve.  Right: Simulated data tends to be much noisier than its experimental counterpart.  Note that as a result, the first local maximum appears in a regime that is most likely part of the linear-elastic regime. }\label{fig:sscurves}
\end{figure}

The discussion in the previous section presupposes a parameterized functional form that can be used to screen and analyze data.  In some cases, however, no such function exists.  This arises, for example, in the case of stress-strain data $\sigma(\epsilon)$, which is used to estimate the yield-strain $\epsilon_y$ of crosslinked polymers (among other materials).  In experimental data, $\epsilon_y$ is typically identified as the first local maximum of $\sigma(\epsilon)$.  Given that such curves are smooth, this point can essentially be identified by eye, so that there is in fact no need to assume a functional form for $\sigma(\epsilon)$.  In simulations, however, thermal noise and finite-size effects can sometimes lead to every other data point being a local maximum, which makes precise estimation of $\epsilon_y$ difficult, to say the least!  See, for example, figure \ref{fig:sscurves}.  Without the benefit of an assumed functional form, we are therefore left to search for more generic mathematical structure that can be used to analyze the data.  This task quickly becomes problem specific, so that we cannot offer general guidelines.  Nonetheless the example of yield-strain illustrates that seemingly obscure mathematical properties can sometimes be of immense help.

Our goal is therefore to state as much as possible about stress-strain curves absent a regression function.  Staring at the left plot of figure \ref{fig:sscurves}, one observes that up to and slightly beyond yield, $\sigma(\epsilon)$ is concave.  Geometrically, this implies that any two points on $\sigma(\epsilon)$ can be connected by a line that lies entirely on or below the curve.\footnote{Analytically this statement is a direct corollary of Jensen's inequality.}  Conceptually, this observation is important because a concave function can have at most one maximum, which, in this case, corresponds to $\epsilon_y$.  Furthermore, we immediately see that the simulated result is manifestly non-concave because noise induces multiple maxima in the data.  This suggests that we could extract $\epsilon_y$ by somehow forcing the data to be concave, or at least interpreting its average behavior as concave.

\begin{figure}
\includegraphics[width=16cm]{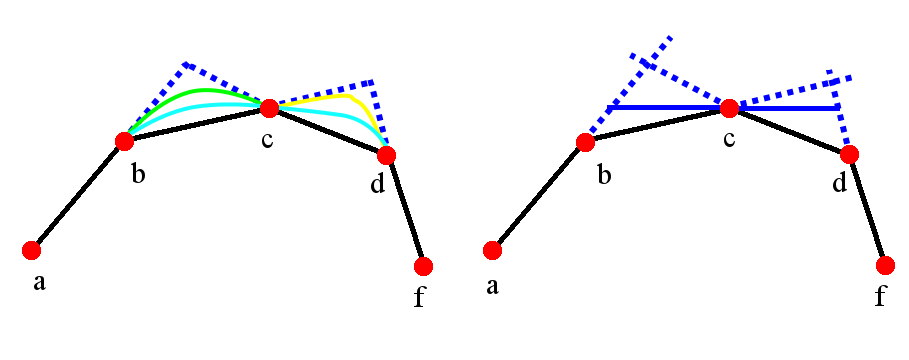}\caption{Illustration of convex fit non-uniqueness. Solid black lines are concave piecewise-linear functions that interpolate points $a$ through $f$ while dotted blue lines extend line segments $ab$, $bc$, $dc$, and $df$.  {\it Left}: The teal curve admits derivatives (slopes) everywhere that are less than the slope of segment $ab$ and greater than the slope of $df$. The green curve has the same slope as segments $ab$ and $cd$ at points $b$ and $c$ respectively. Similarly, the yellow curve has the same slope as segment $bc$ and $df$ at points $c$ and $d$ respectively. Clearly a family of smooth interpolatory curves, all concave, can be constructed.  In fact, it is straightforward to show that the solid-black and dotted-blue lines provide lower and upper bounds on all functions within this family. {\it Right}: The horizontal blue line indicates the domain over which admissible convex functions can have a maximum exceeding that of point $c$; see the left figure for justification.}\label{fig:convices}
\end{figure}

Consistent with our earlier observations, however, there is no unique concave function (e.g.\ in a least-squares sense) that necessarily fits a collection of points.  Figure \ref{fig:convices} illustrates this point.  The different color curves show distinct concave functions that interpolate the same points, labeled a-f.  This family of functions, denoted by $\{g_i\}$, all have the same least-squares residual relative to simulated data, provided the ordinate of the datapoints corresponds to the intersection points of the concave functions.  Notably the black piece-wise interpolation is contained in $\{g_i\}$.  The dotted blue lines are constructed by extending all of the segments of the black curve to their first intersection points (on either side, in the case of more data).  These dotted lines are an upper bound on the admissible values of every curve in $\{g_i\}$, while the black curve is a corresponding lower bound.  This fact can be proved by recourse to geometrical arguments.\footnote{In fact, the essence of the argument is quite simple.  Consider segment bc.  Any curve interpolating bc and falling under the segment is by definition not concave.  Consider, next, the dotted line continuing segment ab.  Assume that a concave curve $\mathfrak g$ both (i) lies above this dotted line, and (ii) interpolates all of the points.  Then one of two possibilities holds.  Either the continuation of $\mathfrak g$ to the left of b lies on or above ab, or it falls below ab.  In the first case, it is obvious that we can find a line connecting two points on $\mathfrak g$ lying above $\mathfrak g$, contrary to assumption.  In the latter case, the segment ab (or some fraction thereof) also lies above the curve, contrary to assumption.  Thus, $\mathfrak g$ cannot lie above the dotted line.  Making these statements more precise requires technical machinery that goes beyond our scope.}

From an UQ standpoint, this non-uniqueness is not problematic and can at times prove useful.  To see this, assume that we have determined a collection of best-fit convex functions $\{g_i\}$ all having the same sum-of-squares relative to the data.  This would correspond, for example, to finding the solid curves in the left plot of figure \ref{fig:convices}.    Then the range of admissible $\epsilon_y$ is given by the set of the corresponding maxima of each $g_i$, namely $\epsilon_y\in \{\epsilon_{i}:\epsilon_i={\rm argmax}_\epsilon \, g_i(\epsilon)\}=: E$.    By the geometrical properties discussed above, it is straightforward to compute this set $E$ exactly.  In particular, denote the set of points common to the family of curves as $(\epsilon_j,\varsigma_j)$ for $1\le j \le N$.  Given that the piece-wise interpolant is a lower bound, while the continuation of piecewise segments is an upper bound (see figure \ref{fig:convices}), one can easily show that $E$ is the closed interval given by $E=[\epsilon_\ell,\epsilon_r]$, where
\begin{align}
\epsilon_\ell &= \epsilon_{j-1} + \frac{\varsigma_j - \varsigma_{j-1}}{m_{j-1}} \label{eq:lowbound}\\
\epsilon_r & = \epsilon_{j+1} + \frac{\varsigma_j - \varsigma_{j+1}}{m_{j+2}} \label{eq:highbound}
\end{align}
if $j={\rm argmax}_k\{\varsigma_k\}$ is unique and $m_k=(\varsigma_k - \varsigma_{k-1})/(\epsilon_k - \epsilon_{k-1})$ is the slope of the lower piecewise interpolant.  This domain is illustrated in the right plot of figure \ref{fig:convices}, which also provides an interpretation of $E$.     If $j={\rm argmax}_k\{\varsigma_k\}$ is not unique, then $\epsilon_\ell = \epsilon_{\min\left[{\rm argmax}_k\{\varsigma_k\} \right]}$ and $\epsilon_r =\epsilon_{ \max\left[{\rm argmax}_k\{\varsigma_k\} \right]}$.  

From a practical standpoint, it is important to note that Eqs.~\eqref{eq:lowbound} and \eqref{eq:highbound} do not actually require that we find the family $\{g_i\}$ of interpolating functions.  Rather, we only need an estimate of their lower bounds, from which we can derive $[\epsilon_{\ell},\epsilon_r]$. 
{\it The task of identifying and bounding the maximum of a stress-strain curve therefore falls to determining the point-wise best-estimates $\varsigma_i$.}

  To accomplish this, first let $\boldsymbol \epsilon = (\epsilon_1,\epsilon_2,...,\epsilon_N)$ denote the vector of $N$ equi-spaced strains sampled by the simulation, with $\boldsymbol \sigma = (\sigma_1,\sigma_2,...,\sigma_N)$  the corresponding stresses.  Next, assume that we have a collection of unknown best-estimates $\boldsymbol \varsigma=(\varsigma_1,\varsigma_2,...,\varsigma_N)$.  We determine the latter by solving the quadratic programming problem
\begin{align}
\boldsymbol \varsigma = \min_{\hvs} \left[\sum_i (\hat \varsigma_i - \sigma_i)^2 \right] \label{eq:qp}
\end{align} 
subject to the linear inequality constraints
\begin{align}
\bA \hvs^{\rm T} &\le {\bf 0}, \label{ineq:amat} \\
\bb_\ell \le & \hvs \le \bb_h \label{ineq:sigbound}
\end{align}
where inequalities are interpreted componentwise, $\bA$ is the finite-difference matrix 
\begin{align}
\bA = \begin{bmatrix}
-2 & 1 & 0 & \dots & 0 \\
1 & -2 & 1 & \ddots & 0 \\
\vdots & \ddots & \ddots & \ddots & \vdots \\
0 & \dots & 1 & -2 & 1 \\
0 & 0 & \dots & 1 & -2
\end{bmatrix}
\end{align}
and $\bb_\ell$, $\bb_h$ are bounds on the admissible values of $\boldsymbol \varsigma$.  Inequality \eqref{ineq:amat} amounts to the requirement that second-order finite differences (which approximate second derivatives) be non-positive, which can be an alternative definition of concave functions, given sufficient smoothness.  The inequality bounds \eqref{ineq:sigbound} characterize the extent to which we believe that the ``true'' underlying convex function deviates from the data.  In this regard, the bounds can be interpreted as estimates of the point-wise noise in the stress data.  

From a practical standpoint, we can define these bounds as follows.  Let $b=p\max_{i}|\sigma_i - \sigma_{i+1}|$ be $p$ times the largest finite difference between any two adjacent stress values, where $p$ is a user defined value.  We can then take $\{\bb_\ell\}_i = \sigma_i - b$ and $\{\bb_h\}_i = \sigma_i + b$, where $\{\bb \}_i$ denotes the $i$th element of the vector.  This definition amounts to the requirement that the ``true'' values $\boldsymbol \varsigma$ cannot deviate from the data by more than $p$-times largest difference between any successive data points.  

\begin{figure}\begin{center}
\includegraphics[width=16cm]{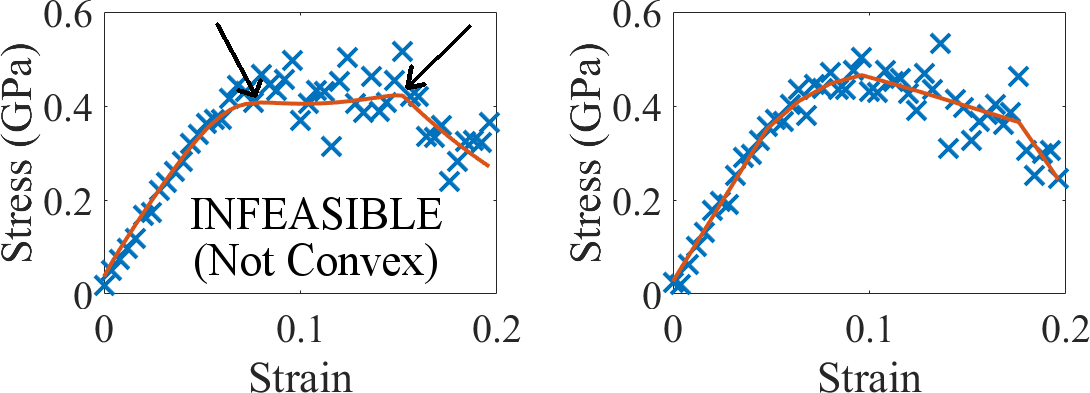}\caption{Attempts to fit simulated data using the QP associated with Eq.~\eqref{eq:qp}.  In all analyses, we choose $p=2/3$ to set a bound on admissible fits; see main text for interpretation of $p$.  Left: A dataset for which the QP is infeasible, meaning that there is not a convex function satisfying the bound constraints.  The solid line is the algorithm's best attempt at a fit (which is not convex).  Although difficult to see, the orange curve has two maxima (indicated by arrows), so that we cannot identify yield on the basis of this computation.  Right: A different dataset for which there is a feasible solution. The points $\varsigma$ have been interpolated with a piece-wise linear curve to yield the solid orange fit.  The fit has a maximum, which can be identified as the yield strain $\epsilon_y$.  }\label{fig:convexfit}
\end{center}
\end{figure}

With the problem now specified by Eqs.~\eqref{eq:qp}--\eqref{ineq:sigbound}, the quadratic program (QP) is either infeasible -- meaning that the constraints cannot be satisfied -- or it returns the optimal values $\varsigma$.  The left plot of figure \ref{fig:convexfit} illustrates this procedure for the data shown in figure \ref{fig:sscurves} taking $p=2/3$, which we find to work well for our collection of datasets.  Note that for this particular dataset, the QP indicates that the problem is infeasible, namely there is no feasible convex function that fits within the noise.  The right plot shows data for which there is a feasible, convex solution.  

For this particular set, it is important to note that the analysis according to Eqs.~\eqref{eq:lowbound}--\eqref{eq:highbound} is not particularly useful since the data is so densely spaced.  This reveals a limitation of using these bounds as the sole basis for uncertainty estimation.  In more detail, we do not in general expect the domain $E$ to be wider than twice the spacing $\Delta \epsilon = \epsilon_j - \epsilon_{j-1}$.  Nonetheless, this does not cause problems {\it per se}, since it means that any estimate of $\epsilon_y$ extracted from Eqs.~\eqref{eq:qp} is relatively well localized.  In such cases, parametric bootstrap and related methods may be useful for estimating uncertainties associated with the fit.  

We end this section by noting that the QP formulation can be modified to address the situation in which one does not know {\it a priori} which subset of the data to treat as convex.  As in the case of hyperbola fits, the goal is to propose a method that is hands-off and avoids potential bias from a modeler.  The key idea is as follows.  For a given collection of stress-strain data, solve the QP problem for the full data.  If the problem is feasible and returns an acceptable estimate of $\epsilon_y$, then stop.  If the problem is infeasible, remove the last data-point from the set, and repeat.  Iterating in this way, one will either (i) arrive at a feasible solution, or (ii) exhaust all of the data (to within a predefined amount).

\begin{figure}
\includegraphics[width=16cm]{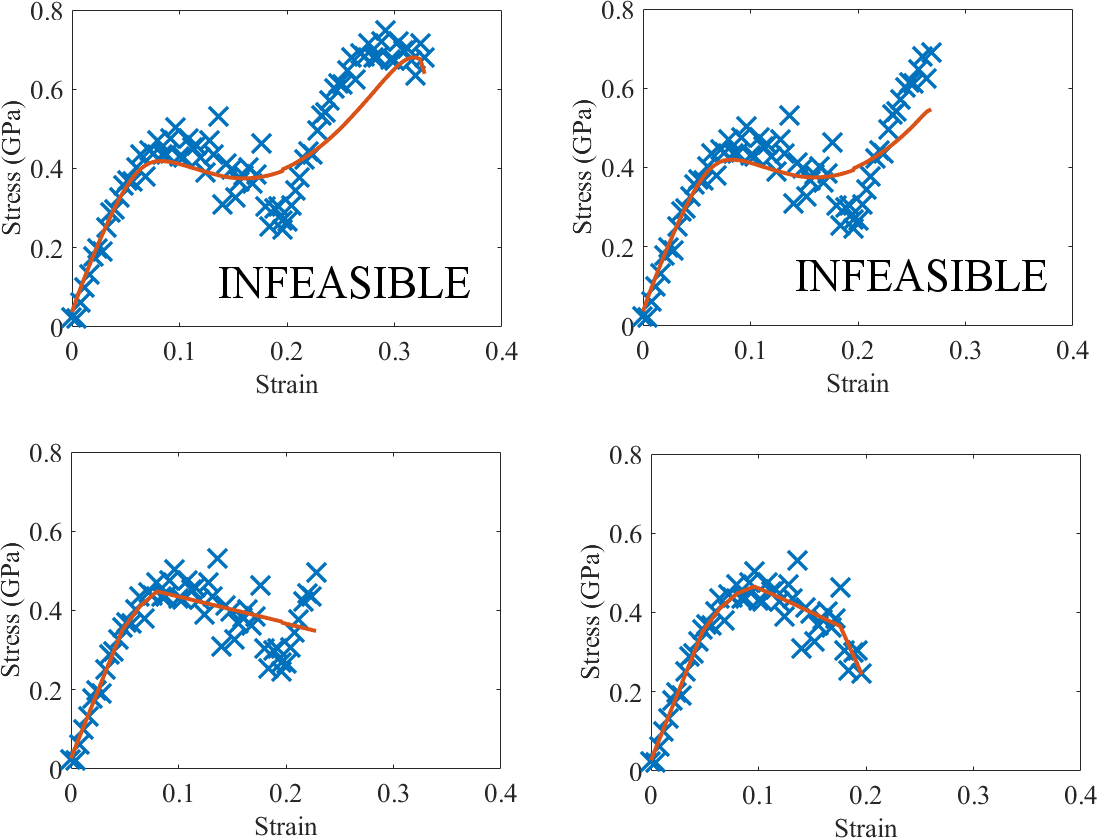}\caption{Analysis of obviously non-convex data in which we iteratively remove data-points.  The top plots show the QP analysis applied to the full data and the effect of removing the last 15 data points.  Solid curves are guides for the eye (output by the algorithm) and are not convex!  According to the analysis, they cannot be used to estimate $\epsilon_y$.  In the bottom left plot, we have removed 40 data points, and the QP now returns a feasible solution.  The right plot shows the result of removing 50 data points, which provides a better fit to the data.}\label{fig:fourplots}
\end{figure}

Figure \ref{fig:fourplots} shows this analysis applied to a somewhat artificial example in which we have duplicated the first 35 datapoints (of a 50 point set) and tacked them on to the end of the stress-strain curve.  We also set $p=2/3$.  The QP returns an infeasible solution for the full data (85 points) and after removing the last 15.  Only after we remove 25 data-points does the QP return a feasible solution.  Admittedly, we do better by removing all 35 of the last data-points, as illustrated by the lower right plot.  Thus, while this method is likely to be useful, it perhaps should not yet be used without some supervision.  

More generally, we emphasize that in the context of this specific examples, many open questions remain.  Efforts to harden these analyses and develop additional uncertainty quantification tools are ongoing.  

\section{Final thoughts}

The examples of the previous sections hopefully serve to illustrate that thoughtful data
analysis is a critical component of any protocol that attempts to extract useful information
from simulations. As a corollary, however, we wish to emphasize that this perspective should
not be limited to simulated data alone. In particular, theory, computation, and experiment
are increasingly being integrated so that hierarchical models inform experiments at various
scales, and vice versa. With this increasing integration comes the reality that uncertainty
and errors will propagate between methods, to the extent that uncertainty propagation is
becoming a key task of any research project. With this in mind, robust data analysis tools
with uncertainty quantification are therefore required for all stages of a workflow, be they
experimental or computational.

{\it Acknowledgments:}  The authors wish to thank Ryan Evans, William Wallace, and Timothy Burns for helpful comments during preparation of this manuscript.  This work is a contribution of the National Institute of Standards and Technology and is not subject to copyright in the United States.

\bibliography{references}

\begin{thebibliography}{10}
\newcommand{\enquote}[1]{``#1''}

\bibitem{ICME1}
{\em Proceedings of the 3rd World Congress on Integrated Computational
  Materials Engineering (ICME)\/}. Wiley, 2015.

\bibitem{IMAJOM}
Dienstfrey, A., Phelan, Frederick~R., J., Christensen, S., Strachan, A.,
  Santosa, F., and Boisvert, R., \enquote{Uncertainty Quantification in
  Materials Modeling,} {\em JOM\/}, Vol.~66, No.~7, 2014, pp.~1342--1344.

\bibitem{economics}
Goldbeck, G., {\em The economic impact of molecular modeling\/}, Goldbeck
  Consulting, 2012.

\bibitem{ASME}
{\em Guide for Verification and Validation in Computational Solid Mechanics\/},
  ASME, 2006.

\bibitem{CalVal1}
McFarland, J. and Mahadevan, S., \enquote{Multivariate significance testing and
  model calibration under uncertainty,} {\em Computer Methods in Applied
  Mechanics and Engineering\/}, Vol.~197, No. 29–32, 2008, pp.~2467 -- 2479,
  Validation Challenge Workshop.

\bibitem{Rizzi1}
Rizzi, F., Jones, R.~E., Debusschere, B.~J., and Knio, O.~M.,
  \enquote{Uncertainty quantification in MD simulations of concentration driven
  ionic flow through a silica nanopore. {I.} {S}ensitivity to physical
  parameters of the pore,} {\em The Journal of Chemical Physics\/}, Vol.~138,
  No.~19, 2013, pp.~194104.

\bibitem{Rizzi2}
Rizzi, F., Jones, R.~E., Debusschere, B.~J., and Knio, O.~M.,
  \enquote{Uncertainty quantification in MD simulations of concentration driven
  ionic flow through a silica nanopore. {II.} {U}ncertain potential
  parameters,} {\em The Journal of Chemical Physics\/}, Vol.~138, No.~19, 2013,
  pp.~194105.

\bibitem{Patrone16}
Patrone, P.~N., Dienstfrey, A., Browning, A.~R., Tucker, S., and Christensen,
  S., \enquote{Uncertainty quantification in molecular dynamics studies of the
  glass transition temperature,} {\em Polymer\/}, Vol.~87, 2016, pp.~246 --
  259.

\bibitem{Patrone17}
Patrone, P.~N., Tucker, S., and Dienstfrey, A., \enquote{Estimating
  yield-strain via deformation-recovery simulations,} {\em Polymer\/},
  Vol.~116, No. Supplement C, 2017, pp.~295 -- 303.

\bibitem{Strachanreview}
Li, C. and Strachan, A., \enquote{Molecular scale simulations on thermoset
  polymers: A review,} {\em Journal of Polymer Science Part B: Polymer
  Physics\/}, Vol.~53, No.~2, 2015, pp.~103--122.

\bibitem{Varshney08}
Varshney, V., Patnaik, S.~S., Roy, A.~K., and Farmer, B.~L., \enquote{A
  Molecular Dynamics Study of Epoxy-Based Networks: Cross-Linking Procedure and
  Prediction of Molecular and Material Properties,} {\em Macromolecules\/},
  Vol.~41, No.~18, 2008, pp.~6837--6842.

\bibitem{Yu09}
Yu, S., Yang, S., and Cho, M., \enquote{Multi-scale modeling of cross-linked
  epoxy nanocomposites,} {\em Polymer\/}, Vol.~50, No.~3, 2009, pp.~945 -- 952.

\bibitem{stevepat1}
Christensen, S. and Senger, J., \enquote{Distortional matrix of epoxy resin and
  diamine,} July~26 2011, US Patent 7,985,808.

\bibitem{stevepat2}
Christensen, S. and Senger, J., \enquote{Distortional matrix of epoxy resin and
  diamine,} June~29 2010, US Patent 7,745,549.

\bibitem{Khare09}
Lin, P.-H. and Khare, R., \enquote{Molecular Simulation of Cross-Linked Epoxy
  and Epoxy−POSS Nanocomposite,} {\em Macromolecules\/}, Vol.~42, No.~12,
  2009, pp.~4319--4327.

\bibitem{Fan07}
Fan, H.~B. and Yuen, M.~M., \enquote{Material properties of the cross-linked
  epoxy resin compound predicted by molecular dynamics simulation,} {\em
  Polymer\/}, Vol.~48, No.~7, 2007, pp.~2174 -- 2178.

\bibitem{Khare12}
Soni, N.~J., Lin, P.-H., and Khare, R., \enquote{Effect of cross-linker length
  on the thermal and volumetric properties of cross-linked epoxy networks: A
  molecular simulation study,} {\em Polymer\/}, Vol.~53, No.~4, 2012, pp.~1015
  -- 1019.

\bibitem{Khare13}
Sirk, T.~W., Khare, K.~S., Karim, M., Lenhart, J.~L., Andzelm, J.~W., McKenna,
  G.~B., and Khare, R., \enquote{High strain rate mechanical properties of a
  cross-linked epoxy across the glass transition,} {\em Polymer\/}, Vol.~54,
  No.~26, 2013, pp.~7048 -- 7057.

\bibitem{Donth}
Donth, E., {\em The Glass Transition: Relaxation Dynamics in Liquids and
  Disordered Materials\/}, Physics and astronomy online library, Springer,
  2001.

\bibitem{smith2013}
Smith, R., {\em Uncertainty Quantification: Theory, Implementation, and
  Applications\/}, Computational Science and Engineering, SIAM, 2013.

\bibitem{RW}
Rasmussen, C. and Williams, C., {\em Gaussian Processes for Machine
  Learning\/}, MIT Press, Cambridge, MA, 2005.

\bibitem{hesterberg:2011}
Hesterberg, T., \enquote{Bootstrap,} {\em Wiley Interdisciplinary Reviews:
  Computational Statistics\/}, Vol.~3, No.~6, 2011, pp.~497--526.

\bibitem{Calibration1}
Wong, R. K.~W., Storlie, C.~B., and Lee, T. C.~M., \enquote{A frequentist
  approach to computer model calibration,} {\em Journal of the Royal
  Statistical Society: Series B (Statistical Methodology)\/}, Vol.~79, No.~2,
  2017, pp.~635--648.

\bibitem{ferson}
Ferson, S. and Siegrist, J., {\em Uncertainty Quantification in Scientific
  Computing\/}, chap. Verified Computation with Probabilities, IFIP Advances in
  Information and Communication Technology, Springer, 2012.

\bibitem{allen89}
Allen, M. and Tildesley, D., {\em Computer Simulation of Liquids\/}, Oxford
  Science Publ, Clarendon Press, 1989.

\bibitem{binder95}
Binder, K., {\em Monte Carlo and Molecular Dynamics Simulations in Polymer
  Science\/}, Oxford University Press, 1995.

\bibitem{Pathria}
Pathria, R.~K., {\em Statistical Mechanics\/}, Butterworth-Heinemann, Oxford,
  2nd ed., 1996.

\bibitem{SalI}
Salacuse, J.~J., Denton, A.~R., and Egelstaff, P.~A., \enquote{Finite-size
  effects in molecular dynamics simulations: Static structure factor and
  compressibility. {I. T}heoretical method,} {\em Phys. Rev. E\/}, Vol.~53,
  1996, pp.~2382--2389.

\bibitem{SalII}
Salacuse, J.~J., Denton, A.~R., Egelstaff, P.~A., Tau, M., and Reatto, L.,
  \enquote{Finite-size effects in molecular dynamics simulations: Static
  structure factor and compressibility. {II. A}pplication to a model krypton
  fluid,} {\em Phys. Rev. E\/}, Vol.~53, 1996, pp.~2390--2401.

\bibitem{Hoover}
Hoover, W.~G., \enquote{Canonical dynamics: Equilibrium phase-space
  distributions,} {\em Phys. Rev. A\/}, Vol.~31, 1985, pp.~1695--1697.

\bibitem{LAMMPS}
Plimpton, S., \enquote{Fast Parallel Algorithms for Short-Range Molecular
  Dynamics,} {\em Journal of Computational Physics\/}, Vol.~117, No.~1, 1995,
  pp.~1 -- 19.

\bibitem{MATLAB}
MATLAB, {\em version 8.1.0 (R2013a)\/}, The MathWorks Inc., Natick,
  Massachusetts, 2013.

\bibitem{PatroneRosch17}
Patrone, P.~N. and Rosch, T.~W., \enquote{Beyond histograms: Efficiently
  estimating radial distribution functions via spectral {M}onte {C}arlo,} {\em
  The Journal of Chemical Physics\/}, Vol.~146, No.~9, 2017, pp.~094107.

\end{thebibliography}
\bibliographystyle{aiaa}
\end{document}